\documentclass[epj]{svjour}
\usepackage{graphicx}
\usepackage{dcolumn}
\usepackage{bm}
\usepackage{here}
\usepackage{color}
\begin{document}
\title{Delocalization in One-Dimensional Tight-Binding Models with Fractal Disorder
}
\author{Hiroaki S. Yamada\inst{1} 
}                     
%
%
\institute{Yamada Physics Research Laboratory, Aoyama 5-7-14-205, Niigata 950-2002, Japan}
%
\date{Received: date / Revised version: date}
%
\abstract{
In the present work, we investigated 
the correlation-induced localization-delocalization transition 
in the one-dimensional tight-binding model with fractal disorder.
We obtained a phase transition diagram from localized to extended states 
based on the normalized localization length by 
controlling the correlation and the disorder strength of the potential. 
In addition, the transition of the diffusive property of wavepacket dynamics 
is shown around the critical point. 
%
} 
\maketitle
\def\ni{\noindent}
\def\nn{\nonumber}
\def\bH{\begin{Huge}}
\def\eH{\end{Huge}}
\def\bL{\begin{Large}}
\def\eL{\end{Large}}
\def\bl{\begin{large}}
\def\el{\end{large}}
\def\beq{\begin{eqnarray}}
\def\eeq{\end{eqnarray}}

\def\eps{\epsilon}
\def\th{\theta}
\def\del{\delta}
\def\omg{\omega}

\def\e{{\rm e}}
\def\exp{{\rm exp}}
\def\arg{{\rm arg}}
\def\Im{{\rm Im}}
\def\Re{{\rm Re}}

\def\sup{\supset}
\def\sub{\subset}
\def\a{\cap}
\def\u{\cup}
\def\bks{\backslash}

\def\ovl{\overline}
\def\unl{\underline}

\def\rar{\rightarrow}
\def\Rar{\Rightarrow}
\def\lar{\leftarrow}
\def\Lar{\Leftarrow}
\def\bar{\leftrightarrow}
\def\Bar{\Leftrightarrow}

\def\pr{\partial}

\def\Bstar{\bL $\star$ \eL}

\def\etath{\eta_{th}}
\def\irrev{{\mathcal R}}
\def\e{{\rm e}}
\def\noise{n}
\def\hatp{\hat{p}}
\def\hatq{\hat{q}}
\def\hatU{\hat{U}}

\def\iset{\mathcal{I}}
\def\fset{\mathcal{F}}
\def\pr{\partial}
\def\traj{\ell}
\def\eps{\epsilon}
\section{Introduction}
In some kinds of one-dimensional Schr$\ddot{o}$dinger operators with random potential,
the eigenstates  are mathematically proven to be localized \cite{ishii73,lifshiz88,stollmann01}.
The one-dimensional tight-binding models with an ergodic
and stationary random potential have
 positive Lyapunov exponent 
of the wavefunction with probability 1 (G-M-P theorem) \cite{goldsheid77}.
The existence of the positive Lyapunov exponent is necessary and sufficient condition
for a pure point set spectrum of the operators, 
and then all the eigenfunctions exhhibit the exponentially decay  in the thermodynamic limit.
Kotani's theory states that 
if the potential sequence is nondeterministic under the following conditions, 
(i)stationarity,
(ii)ergodicity,
(iii) integrability,
then there is no absolutely continuous (a.c.) spectrum of the operators \cite{kotani82}.
These theorems can be proven true for continuous and discrete one-dimensional disordered 
systems (1DDS) \cite{simon83,kotani86,damanik04}.
However, the necessary and sufficient condition for the exponential
localization has not been found yet.

There is a possibility that the correlation effect of the potential sequence 
breaks the strong exponential localization and generates 
a localization-delocalization transition (LDT) in the 1DDS.
Indeed, many authors numerically observed 
the correlation-induced LDT by using the some 
 potential sequences with power spectrum
$S(f) \sim 1/f^\alpha$($\alpha \geq 2$) as a potential 
by Fourier filtering method (FFM),  where $f$ denotes frequency 
\cite{moura98,izrailev99,gong10,deng12,kaya07,gong12,albrecht12}. 
Such a  potential obviously breaks a necessary condition (i), i.e. it is nonstationary.
Note that the non-stationarity can be satisfied for $\alpha \geq 1$.
In addition,  the LDT with mobility edges, has been  found numerically for 
the one-dimensional tight-binding model with 
the FFM model
as the potential strength $W$ decreases 
\cite{bocker05,oliveira01,shima04,greis81,garcia09,garcia10,costa11}.

Shima {\it et al} and Kaya also showed that the 1DDS with FFM potential
have a critical disorder strength $W_c$ separating the conducting and insulating
phases,  and the  $W_c$ is independent of the spectrum index $\alpha( >2)$
 \cite{shima04,kaya07}. 
Petercen and Sandler insisted that effect of the anticorrelation is also 
important to understand the transition due to the correlation 
of the sequence \cite{petersen13}.

On the other hand, the LDT due to 
the differentiability of the potential function 
 also exists without contradiction with the Kotani's theory.
Very recently, Garcia and Cuevas studied the transition
 based on the differentiability of 
the disorder potential as a necessary condition 
for the delocalization \cite{garcia09,garcia10}. 
A certain degree of differentiability assures that the potential 
in neighboring sites is strongly correlated.
They modeled the sequences with power-law spectrum by Weierstrass function 
with fractal dimension $D$. 
As a result they also numerically suggested that the transition 
takes place at the critical value $D_c=3/2$
by means of the distribution of the energy level-spacing
in the weak disorder limit. 
However, studying the effect of the disorder
strength in the 1DDS with the Weierstrass-type potential has been still unclarified.

In this paper, we have numerically studied the correlation-induced
 localization-delocalization transition 
by using tight-binding model with Weierstrass potential
 used by Garcia and Guevas \cite{garcia10}.
 The finite-size scaling analysis 
for the normalized localization length 
at band center numerically suggests the existence 
of the transition around $D \simeq 3/2$ independent of the potential strength
in the  relatively weak disorder regime.
On the other hand, in the relatively strong disorder regime,
the critical fractal dimension $D_c$ becomes smaller value than $D=3/2$ dependently on the 
potential strength.
Furthermore, we have investigated the quantum diffusion of the initially localized 
wavepacket in the system.  The transition from the localized state 
to ballistic states occurs 
around $D\simeq 3/2$ without scale invariant subdiffusive behavior. 

This paper is organized as follows.
In the next section, we introduce the 1DDS with the Weierstrass potential 
and some preliminary calculations.
In Sect.\ref{sect:main}, we present 
global behavior of the $W-$dependence and $D-$dependence
in the LDT by the numerical calculation of the normalized localization length.
 A phase diagram in the $D-W$ space is also given.
In Sect.\ref{sect:diffusion}, we show that the dynamical property of the system
 ought to be represented by the time dependence of the degree 
to which the initially localized wavepacket would be spread.
Summary and discussion  are presented in the last section.
In an appendix, we give the behavior of the autocorrelation functions for the 
potential sequence.

\section{Model and preliminary calculation}
\label{sect:model}

\subsection{model}
We consider the one-dimensional tight-binding model describing 
single-particle electronic states  
in the site representation as
\begin{eqnarray}
u(n-1)+u(n+1)+ WV(n) u(n) =E u(n),
\label{eq:tight-binding}
\end{eqnarray}
where $E$ and  $\{ u(n) \}_{n=0}^{N}$ are the energy and 
 state of the system,  respectively.  
The $\{ V(n) \}_{n=0}^{N}$  and $W$ are the on-site energy sequence and 
the  strength, respectively.
To model the correlated and non-differential disorder potential 
for $V(n)$($n\leq N$) in Eq.(\ref{eq:tight-binding}), 
we use the following form:
\beq
V(n) = C \sum_{k=0}^{L} \frac{\sin(2\pi a^k n/N +\phi_k)}{a^{(2-D)k} },
\label{eq:wei-pot}
\eeq
where $a$ is a constant value ($a>1$) related the scale-invariance 
and $D$ is a fractal dimension ($1<D<2$).
 $\{ \phi_k \}_{k=0}^L$ are random independent variables chosen in the interval $[0,2\pi]$.
$C$ is the normalization constant which is determined 
by a condition 
\beq
\sqrt{<V(n)^2>-<V(n)>^2}=1,
\eeq
where $\langle...\rangle$ indicates the average over realization
of the phases in Eq.(\ref{eq:wei-pot}).

If we set $n/N=x$, $\phi_k=0$,
the potential sequence becomes "Weierstrass function" with continuous and  
indifferentiable everywhere by taking 
a continuous limit $N \to \infty$ and $L \to \infty$. 
Therefore, the potential will be shortly transfered to as "Weierstrass potential" in this paper, and 
we set $a=2$ and $L=50$ thorough this paper 
without loss of the generality and accuracy of the numerical calculation.
The power spectrum $S(f)$ of the Weierstrass function 
is empirically characterized by the fractal dimension $D$ as,
\beq
S(f) \sim \frac{1}{f^{5-2D}}.
\eeq
In compassion with the form $S(f)\sim 1/f^{\alpha }$, 
\beq
D = 1+ \frac{3-\alpha}{2}.
\eeq
Note that the 
condition $\alpha \geq 2$ for the LDT
in the FFM potential \cite{moura98,shima04,kaya07} 
corresponds to a condition $ D \leq 3/2$.
Increasing $\alpha$ corresponds to increasing correlations up to 
long-range correlated disorder.
The analytical property of the autocorrelation function for 
the Weierstrass potential $V(n)$
is given in appendix \ref{app:correlation}. 
It is found that the correlation linearly decreases 
from 1 for $D=3/2$.

In addition to the long-rang correlation, the fractal dimension $D$ 
also controls the degree of the differentiability of the potential function.
The degree of the differentiability increases along with the decrease of the fractal
dimension $D$.
The smoothness of the potential fluctuation can also induce the delocalization 
of the quantum states, which property is directly related to analyticity of the 
potential function in the continuum limit, as pointed out by Garcia and Cuevas 
 \cite{garcia09,garcia10}.
They have numerically found that LDT at $D_c=3/2$ for the sufficiently weak 
disorder regime  by using the nearest-neighbor level-space distribution 
of the energy spectrum.
The result is not at odds with Kotani's theory because 
the Weierstrass potential become non-stationary for $1 < D \leq 3/2$.

\subsection{preliminary calculation}
\label{sect:preliminary}
The finite size Lyapunov exponent $\gamma_N (N>>1)$ of 
the one-dimensional systems can be defined by 
\begin{eqnarray}
\gamma_N &=&  \frac{ \ln \left( |u(N)|^2 + |u(N+1)|^2 \right) }{2N},
\label{eq:gamma_n}
\end{eqnarray}
with initial state $u(0)=u(1)=1$. 
Then the Lyapunov exponent $\gamma$ and the localization length $\xi$ are
 given by $\gamma = \lim_{n \to \infty} \gamma_N$ and by $\xi=1/\gamma$, respectively.
The energy dependence of  $\gamma$ is strongly correlated with the density of states
that can be obtained from some experiments for real materials.
In addition, we define the normalized localization length (NLL), 
\beq
\Lambda_N \equiv \frac{\xi(N)}{N}=\frac{1}{<\gamma_N> N},  
\eeq
, where $\langle...\rangle$ denotes the ensemble average.
It is useful to study the LDT that $\Lambda_N$ decreases (increases) with the system size 
$N$ for localized (extended) states,
and it becomes constant for the critical states.

\begin{figure}[htbp]
\begin{center}
\includegraphics[width=8.6cm]{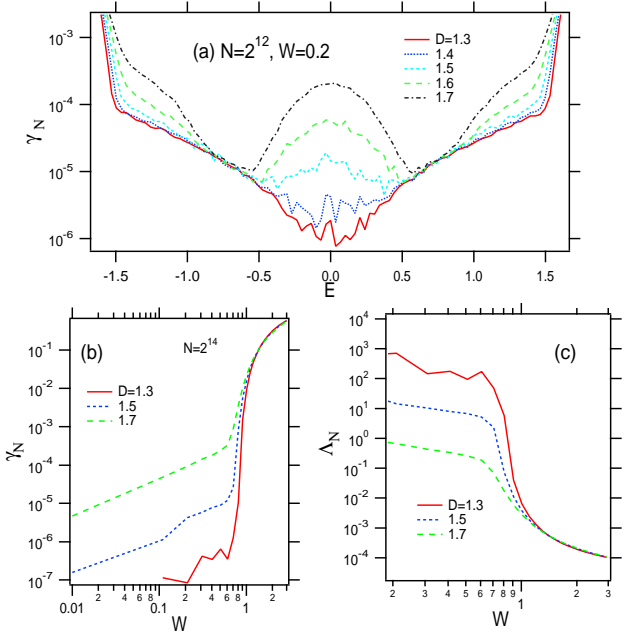}
\caption{(Color online)
(a)The Lyapunov exponent $\gamma_N$ as a function of energy $E$ 
for a weak disorder case, $W=0.2$ and varying the fractal dimension $D$.
The system and ensemble sizes are $N=2^{12}$ and $2^{10}$, respectively.
(b)The Lyapunov exponent $\gamma_N$ 
and 
(c)normalized localization length $\Lambda_N$ at the band center $E=0$
as a function of the potential strength $W$
for the several values of the $D$. 
This system and ensemble sizes are $N=2^{14}$ and $2^{10}$, respectively.
}
\label{fig:energy-fig1}
\end{center}
\end{figure}

 Figure \ref{fig:energy-fig1}(a) shows the energy dependence of the
Lyapunov exponent $\gamma_N$ for some values of the fractal dimension $D$.
The Lyapunov exponent $\gamma_N$ at the band center $E=0$ decreases
as the value of the $D$ decreases.
It suggests a possibility of the delocalized states (extended states) 
at the band center $E=0$ for small values of $D(\leq 3/2)$.
Figure \ref{fig:energy-fig1}(b) and (c) 
show the potential strength dependence of 
the Lyapunov exponent $<\gamma_N>$ and the renormalized localization length $\Lambda_N$
at the band center $E=0$ for some parameter sets.
In the weak disorder regime $W<<1$, the numerical data lead to 
\beq
  \gamma_N \propto W,
\label{eq:gamma-w^2}
\eeq
despite of the system size, which is equivalent to case of the uncorrelated 
disordered system. In the strong disorder regime, the Lyapunov exponent does not 
depend on the system size $N$.
Figure \ref{fig:energy-fig1}(c) shows the $W-$dependence of 
the NLL $\Lambda_N$.
It is found that 
the localization length is larger than system size in the week disorder limit.
The results suggest that for the strongly correlated limit, 
there is a possibility of the states with $<\gamma_N> \to 0$ and/or $\Lambda_N >1 $
in the thermodynamic limit $N \to \infty$, 
which correspond to a delocalized phase with  the extended states
in week disorder limit.
In the following section, we numerically 
investigate the $D-$dependence, $W-$dependence 
and $N-$dependence of the NLL 
in the 1DDS with the potential (\ref{eq:wei-pot}).

\section{Localization-Delocalization Transition}
\label{sect:main}
This is the main section of the present paper.
In what follows, we investigate the NLL 
at band center $E=0$  by changing the system size 
for some typical parameter sets ($W$,$D$).
The typical basis size $N$ and ensemble size used 
here are $N=2^{8} \sim 2^{17}$ and $2^{10}\sim 2^{12}$, respectively.
The robustness of the numerical 
calculations with respect to the system size has been confirmed in each case.

\subsection{$D$-dependence and $W$-dependence
 in the wide parameters space}

\begin{figure}[htbp]
\begin{center}
\includegraphics[width=8.5cm]{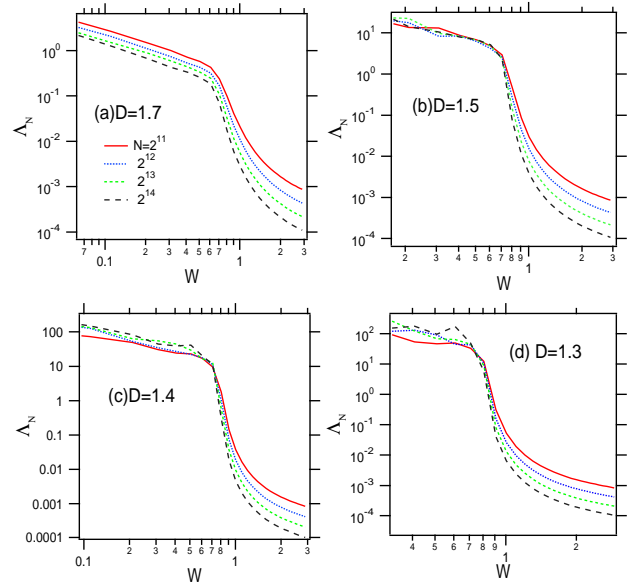}
\caption{(Color online)
The normalized localization length $\Lambda_N$ as a function of 
the potential strength $W$ with a fixed value of D for  
 $N=2^{11},2^{12},2^{13},2^{14},2^{15},2^{16}$.
 (a)$D=1.7$, (b)$D=1.5$, (c)$D=1.4$, (d)$D=1.3$.
Note that the data are plotted in double-logarithmic scale.
}
\label{fig:fig3-w-dep}
\end{center}
\end{figure}

\begin{figure}[htbp]
\begin{center}
\includegraphics[width=9.3cm]{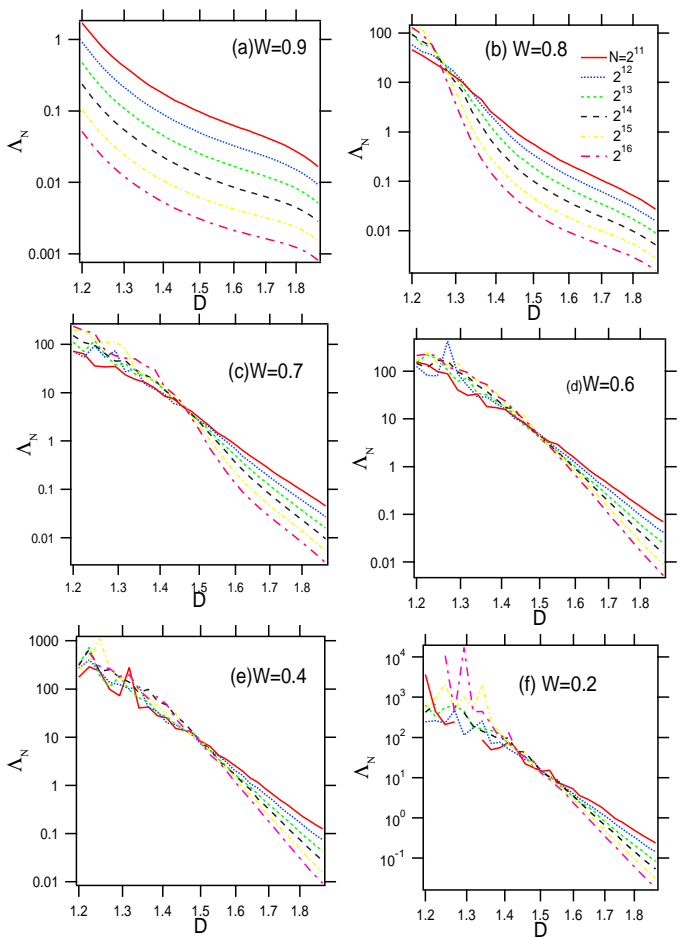}
\caption{(Color online)
The normalized localization length $\Lambda_N$ as a function of $D$
with a fixed value of W for  $N=2^{11},2^{12},2^{13},2^{14},2^{15},2^{16}$.
(a)$W=0.9$, (b)$W=0.8$, (c)$W=0.7$, (d)$W=0.6$, (e)$W=0.4$, (f)$W=0.2$.
Note that the vertical axis are in logarithmic scale.
}
\label{fig:fig3-main}
\end{center}
\end{figure}

Figure \ref{fig:fig3-w-dep} shows $W-$dependence of the NLL 
for some values of the $D$  in the relatively wide range.
The $W-$dependence relatively smoothly drops down around $W \sim 1$ 
in the same way for all cases with different system size. 
There exists the strong system size dependence for the strong disorder regime $W>1$.
Furthermore it is found that there is 
a quite different feature between the cases for $D<3/2$ and ones for $D>3/2$.
In the case of $D=1.7$, the curve of the $W-$dependence for the different 
system size ($N=2^{11}\sim 2^{14}$) do not cross 
within this regime of  $W$ with each others.
This fact implies that all states around $E=0$ are localized in the thermodynamic limit 
and there is no transition for changing the value of $W$.
On the other hand, in the cases of $D \leq  3/2$, 
the system size dependence for weak disorder regime $W<1$ becomes very weak,  and 
the $W-$dependence sharply decreases at certain value of $W$.
Apparently, we can expect that for $D \leq 3/2$ the $W-$dependence of the NLL 
shows a sharp jump in the thermodynamic limit $N \to \infty$.
This feature suggests the existence of a transition to delocalized states
in the limit $N \to \infty$.

In Fig.\ref{fig:fig3-main} we show the $D-$dependence of the NLL
for some values of $W$ in order to find out the critical value $D_c$ that
all curves of different system size intersect at the value.  
At least, the existence of the intersection corresponding to a transition point 
can be observed around $D=3/2$ for the relatively weak disorder strength $W \leq 0.7$.

In the following two subsections, we investigate the details of the LDT 
with focusing on the relatively weak disorder regime ($W \leq 0.7$) and 
the relatively strong disorder regime ($0.7<W<0.8$), respectively.


\subsection{Weak disorder regime}
\begin{figure}[htbp]
\begin{center}
\includegraphics[width=6.5cm]{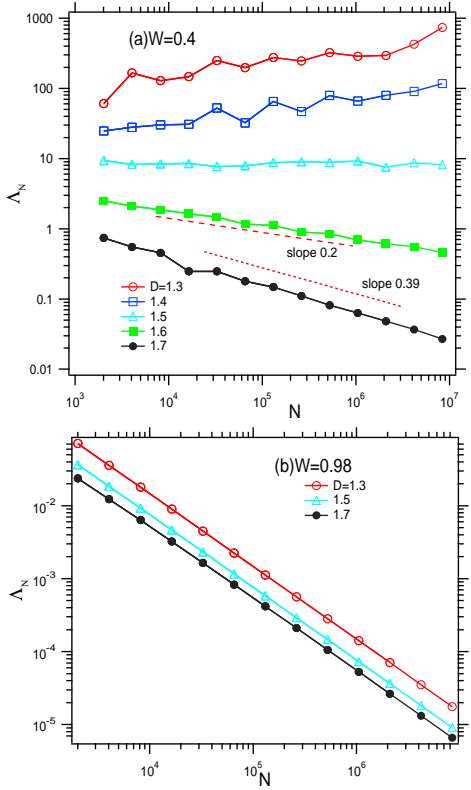}
\caption{
(Color online)
The normalized localization length $\Lambda_N$ as a function of the 
system size $N$ with several values of the fractal dimension
for the potential strength, (a)$W=0.4$ and (b)$W=0.98$.
Note that the data plotted in double-logarithmic scale.
}
\label{fig:n-dep-loclength}
\end{center}
\end{figure}

\begin{figure}[htbp]
\begin{center}
\includegraphics[width=6.5cm]{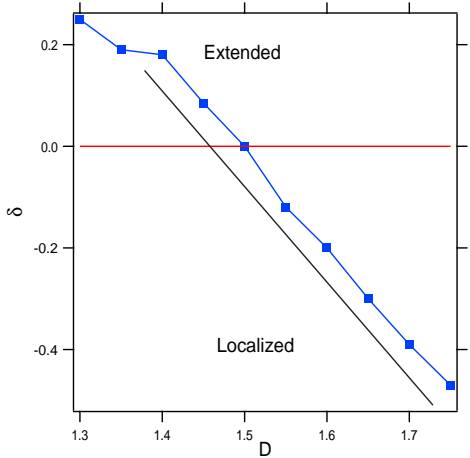}
\caption{
(Color online)
The index $\delta$ of Eq.(\ref{eq:delta}) as a function of the 
fractal dimension $D$ for the case $W=0.4$ in Fig.\ref{fig:n-dep-loclength}. 
The result has been obtained by the least square fit for the $N-$dependence 
of the normalized localization length $\Lambda_N$.
The straight line with a slope $-1.88$ is shown as a guide for eye. 
}
\label{fig:delta}
\end{center}
\end{figure}

Figure \ref{fig:n-dep-loclength}(a) shows the system size dependence 
of the NLL for a relatively weak disorder strength $W=0.4$.
It is found that the $N-$dependence changes the decreasing function
 to the increasing function as the fractal dimension decreases, and the 
 $N-$dependences are algebraic.
Generally, the quantum states can be classified by the exponent $\delta$ of the
 $N-$dependence of the NLL $\Lambda_N$ 
when it  behaves as,
\beq
 \Lambda_N \sim N^{\delta}.
\label{eq:delta}
\eeq
The exponent, $\delta < 0$ for the localized states,
$\delta > 0$ for the extended states, and 
$\delta = 0$ for the critical states.
The exponents $\delta$ obtained by the least-square method are shown 
in Fig.\ref{fig:delta}. 
The exponent $\delta$ decreases with respect to the fractal dimension $D$ as
\beq
 \delta \sim (3/2-D)^{-1.88}.
\eeq
It seems that in the $N-$dependence of the NLL 
the sign of the index $\delta$ changes from negative to positive one 
at the point $D_c=3/2$ for $W=0.4$ in Fig.\ref{fig:n-dep-loclength}(a).

As a result  it is suggested that 
the LDT takes place around the transition point $D_c=3/2$ independent of the disorder strength
in relatively weak disorder regime $W <0.7$, 
as shown in Fig.\ref{fig:fig3-w-dep}, Fig.\ref{fig:fig3-main}, and Fig.\ref{fig:n-dep-loclength}.
 
However,  it should be noticed that the transition does not obey 
the standard one-parameter scaling theory (OPST) 
of the localization \cite{abrahams79}
if  the $N-$dependence seen in Fig.\ref{fig:n-dep-loclength}(a)
continues in the thermodynamic limit $N \to \infty$.
In the present stage, it is difficut to get  the $N-$dependence with adequate accuracy 
for the larger system size because of the limitation of the numerical calculation.

\subsection{Strong disorder regime}
We examine the  $\Lambda_N$ behavior
to get the delocalization  in the relatively strong disorder regime ($W >0.7 $).

Figure \ref{fig:n-dep-loclength}(b) shows the system size dependence 
of the NLL for $W=0.98$.
It clearly shows that 
the wavefunction goes to localized states in the thermodynamic limit 
as $\Lambda_N \sim N^{-1}$ irrespective of the fractal dimension
for the relatively large disorder strength $W=0.98$.
Surely we have to investigate the smaller values of $D(\sim 1)$ 
to make the delocalized states for $W=0.98$ although it is very hard.

\begin{figure}[htbp]
\begin{center}
\includegraphics[width=8.5cm]{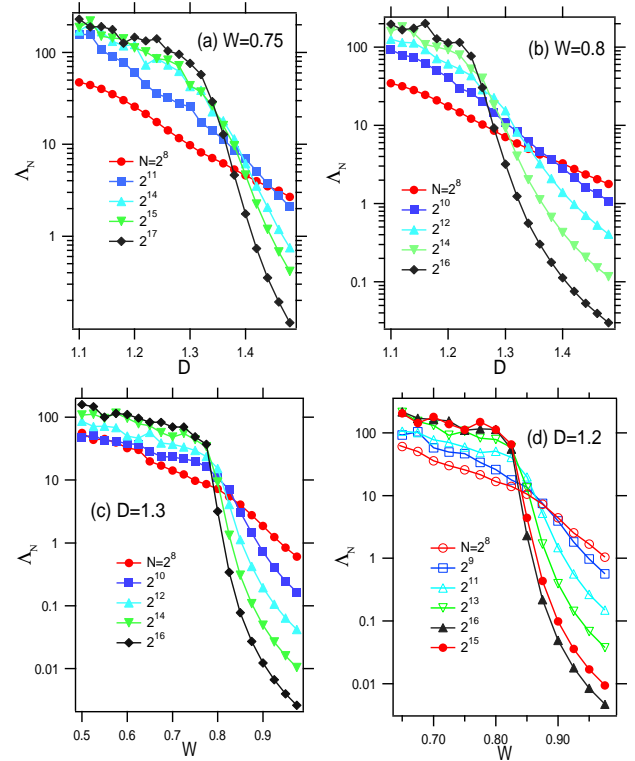}
\caption{(Color online)
Some enlargements of the $D-$dependence and $W-$dependence 
of the normalized localization length $\Lambda_N$.
 (a)$W=0.75$ and (b)$W=0.8$,  (c)$D=1.2$, (d)$D=1.3$.
The system size $N$ is varied as $2^8$ $\sim 2^{17}$.
The averaging is taken over $2^{12}$ realizations.
Note that the data are plotted in the semi-logarithmic scale. 
}
\label{fig:figA}
\end{center}
\end{figure}

Figure \ref{fig:figA} illustrates the more detailed behavior of   $\Lambda_N$ 
as a function of $D$ and $W$ for $D<3/2$ and the relatively strong 
disorder $W>0.7$.
It is found that the crossing points gradually shift and converge to certain value 
in the thermodynamic limit $N \to \infty$.
This result suggests that 
the critical value of $D$ decreases less than $3/2$ 
for the relatively large potential strength $W$. 

\begin{figure}[htbp]
\begin{center}
\includegraphics[width=8cm]{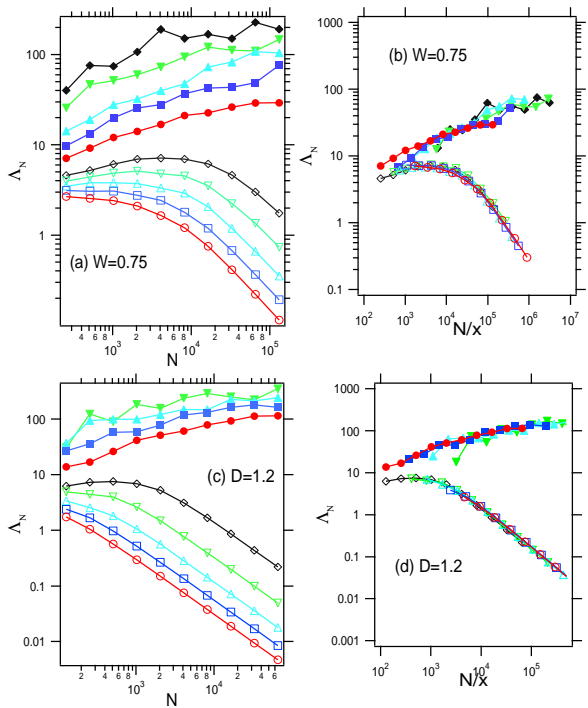}
\caption{
(Color online)
The normalized localization length $\Lambda_N$ as a function of the 
system size $N$ with (a)$W=0.75$ for the different values of  $D$.
$\Lambda_N$ as a function of the  system size $N$ with (c)$D=1.2$
for the values of $W$. 
Panels  (b) and (d) are the scaling function construction from the numerical 
data in panels (a) and (c), respectively.
Note that all are plotted in double-logarithmic scales.
}
\label{fig:figB}
\end{center}
\end{figure}

The $N-$dependence of the NLL for Fig.\ref{fig:figA}(a) and (d) 
 are drawn in Fig.\ref{fig:figB}(a) and (c).
The latter both reveal a clear change  from the decreasing function to  increasing one,  
as the $D$ and $W$ decreases, respectively.
Next, we try to construct the scaling function by the parallel shift 
of the horizontal axis.
If  the OPST around the LDT  exists, then the NLL behaviors as 
\beq
   \Lambda_N=f(N/\xi),
\eeq
where  $f(...)$ is the scaling function and  $\xi $ is the amount of the 
parallel shift,  corresponding to the localization length or correlation length.
In Fig.\ref{fig:figB}(b) and (d), we show the result for the latter shift 
in the Fig.\ref{fig:figB}(a) and (c), respectively.
The  general forms of the graph are similar to the results 
for 1DDS with FFM-potential \cite{shima04}.

The upper (lower) brunches are delocalized (localized) regimes.
As seen in Fig.\ref{fig:figB}(b) and (d),  
it seems  that the data of the lower branches look like being on a common  curve
while the data for the upper branches do not belong to an universal curve, 
as far as present data show.
The $N-$dependence of  $\Lambda_N$ with adequate numerical
accuracy for the larger system size is necessary 
 to perform the finite-size scaling analysis around the critical point.

\subsection{phase diagram}

\begin{figure}[htbp]
\begin{center}
\includegraphics[width=9cm]{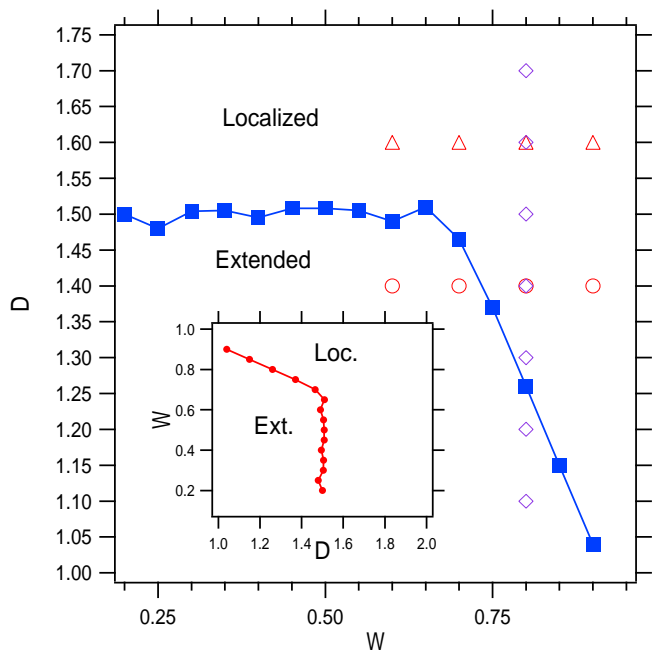}
\caption{
\label{fig:Lyap-fig1}(Color online)
Phase diagram of the localization-delocalization
transition in the $W-D$ plane.
The transition points (denoted by the filled blue squares) have been obtained by 
numerical plots of the normalized localization length as shown in 
Fig.\ref{fig:fig3-w-dep}, Fig.\ref{fig:fig3-main} and Fig.\ref{fig:figA}.
The inset shows the diagram in the $D-W$ plane.
The open symbols (circles, triangles, rhombuses) indicate the parameter sets
used in the quantum diffusion in Fig.\ref{fig:msd-1} and Fig.\ref{fig:msd-2}.
}
\label{fig:phase-diagram}
\end{center}
\end{figure}

We determined the critical values of $D$ and/or $W$ 
by the $D-$dependence and $W-$dependence of 
NLL as seen in Fig.\ref{fig:fig3-w-dep}, Fig.\ref{fig:fig3-main}, 

Fig.\ref{fig:n-dep-loclength} andFig.\ref{fig:figA}. 
Figure \ref{fig:phase-diagram} shows the phase diagram 
in the $D-W$ space separating the localized and delocalized states.
The most interesting point here is that the critical disorder strength $W_c$
depends on $D$ in the strong disorder regime   ($W > 0.7$).
This result implies that the LDT with the $D-$dependent  critical 
disorder strength is  different from the one observed in the 1DDS with FFM-potential, 
 in which  it is $W_c=2$ in our notation
independent of  the spectrum exponent $\alpha$ \cite{shima04}.
And our result  in part coincides with one for the FFM-system by Kaya 
that the critical values of $W$ increases when $\alpha$ increases, 
while the critical exponent $\nu$ decreases when $\alpha$ increases \cite{kaya07}.
Accordingly this result suggests that the phase diagram in the $\alpha-W$ space 
might be different 
even if the potential sequences are 
characterized by the same exponent $\alpha$ of the power spectrum 
in the LDT in the 1DDS.

\section{Quantum diffusion}
\label{sect:diffusion}
It can be expected that for sufficiently differentiable potentials a band 
of delocalized states occurs due to destruction of the interference effects in the 
reflected components of the wavepacket.

In this section, we examine the quantum diffusion 
of the initially localized wave packet by changing the 
parameter $D$.
We monitor the mean square displacement (MSD) of the wavepacket,
\begin{eqnarray}
m_2(t) = \sum_{n}(n-n_0)^2 |u(n,t)|^2,
\end{eqnarray}
where $n_0$ is an initially localized site.
The quantum time-evolution is given by,
\beq
i\hbar \frac{\pr u(n,t)}{\pr t}= u(n+1,t)+ u(n-1,t) + WV(n) u(n,t),
\eeq
where $ n=1,2,...,N$ with initial state $u(n,t=0)=\delta_{n,n_0}$ 
and $\hbar=1$.

\begin{figure}[htbp]
\begin{center}
\includegraphics[width=8.5cm]{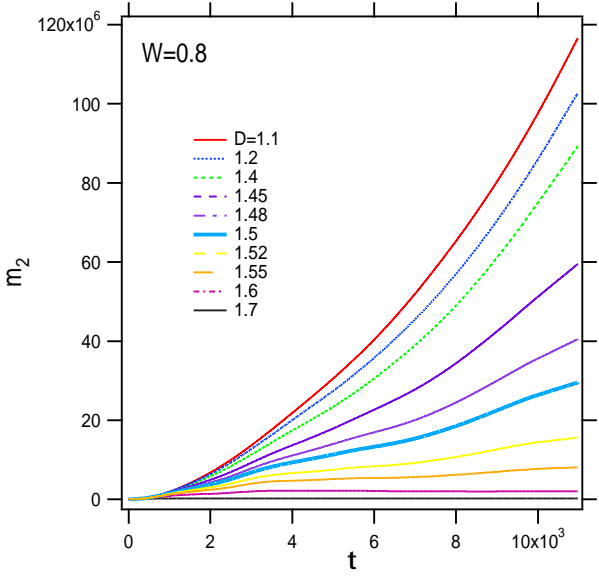}
\caption{
\label{fig:Lyap-fig1}(Color online)
The second moment $m_2$ as a function of time for some fractal dimensions
with $W=0.8$.
The bold curve corresponds to the critical case $D=3/2$.
We set $\hbar=1$ and $\delta t=0.05$. 
The system size and sample sizes are $N=2^{16}$ and $10$, respectively.
}
\label{fig:msd-1}
\end{center}
\end{figure}

Figure \ref{fig:msd-1} shows the time-dependence of the MSD 
for a fixed potential strength $W=0.8$ and varying the fractal dimension $D$.
The parameter sets we used are denoted in Fig.\ref{fig:phase-diagram} by 
some open symbols.
Apparently, the $m_2$ ballistically grows for the relatively small values of $D$, 
and it  is  localized for the cases with $D>3/2$.

\begin{figure}[htbp]
\begin{center}
\includegraphics[width=7cm]{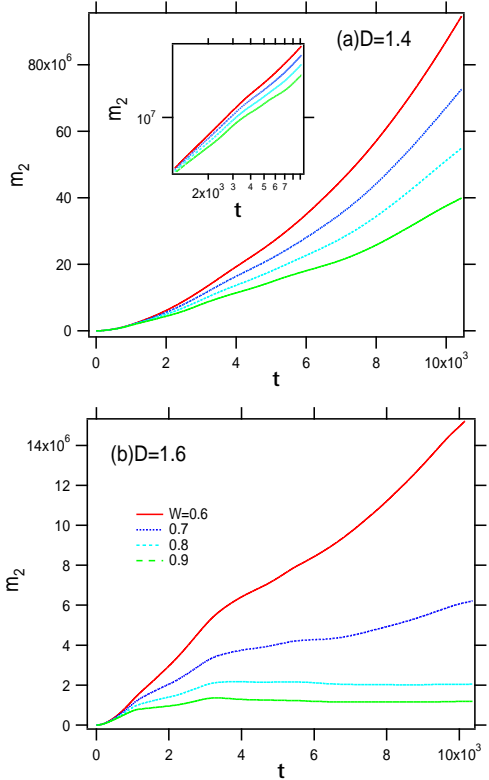}
\caption{
\label{fig:Lyap-fig1}(Color online)
The second moment $m_2$ as a function of time for some values of 
$W$ ($W=0.6, 0.7, 0.8, 0.9$) at (a)$D=1.6$ and (b)$D=1.4$.
We set $\hbar=1$ and $\delta t=0.05$. 
The system size and sample sizes are $N=2^{16}$ and $10$, respectively.
The inset shows the log-log plots.
}
\label{fig:msd-2}
\end{center}
\end{figure}

Next, Fig.\ref{fig:msd-2} shows the diffusive properties for some values of  
the potential strength with fractal dimension, $D=1.4$, $D=1.6$,
denoted by the open symbols in Fig.\ref{fig:phase-diagram}.
It follows that the ballistic-like motion ($m_2 \sim t^{\beta}, \beta >1$)
can be obtained for all cases with $D=1.4$, while it is well-localized
for the relatively larger value of $W$ in the cases with $D=1.6$.
The all of the cases of $D=1.6$ are localized 
for the long-time calculation. (It is not shown here.)

As a result, it seems that in the quantum diffusion 
the critical value $D_c$ also around $D \sim 3/2$
which is consistent with the result by the normalized localization length
$\Lambda_N$ in the last section.
 However, clear subdiffusive behavior have not been observed around the 
critical point $D_c=3/2$, which is different from 
the localization-delocalization transition
in three-dimensional disordered systems and 
polychromatically perturbed 1DDS, 
in which cases an asymptotic behavior, $m_2 \sim t^{2/3}$, has been obtained 
at the critical point
\cite{yamada99,yamada04,delande08}.
This discrepancy is not so surprising if  
the correlation-induced LDT deviates from the standard OPST
as shown in  the previous section.

\section{Summary and Discussion}
\label{sect:summary}
In this work we numerically investigated the combined effects of the disorder
strength  $W$ and the order of the long-range correlation 
characterized by fractal dimension $D$ 
in the one-dimensional tight-binding model with the Weierstrass potential.
The result based on the normalized localization length 
strongly suggests that quantum states are localized for $D >3/2$,
whereas  we have obtained critical
disorder strength $W_c$ which separates extended and localized
regimes for $D \leq 3/2$.
 In particular,  
the result that the critical value $D_c=3/2$ is not depend on 
the $W$ is consistent with the result given by Garcia and Cuevas
in the weak disorder limit.
While the critical value of the fractal dimension
depends on the potential strength in the strong disorder regime.
Furthermore, the localization-delocalization property reflected on the 
quantum diffusion of the initially delocalized wavepacket 
although the subdiffusion could not be observed at the critical case.
The disorder strength dependence of the correlation-induced transition 
is one of the interesting features in the  localization-delocalization transition.

Concerning the power spectrum $S(f) \sim 1/f^\alpha, \alpha=2$,
the localization-delocalization transition point $D_c=3/2$ is consistent 
with one predicted in the system with some other correlated potential 
such as FFM potential \cite{izrailev99,esmailpour07,pinto04,croy11,petersen12}.
Moreover, there are the other types of 
long-range correlated potential with discrete values 
such as binary and ternary sequences \cite{pinto04,yamada91}.
One of the common points between the discrete and continuous models 
is nonstationary of the potential sequences caused by the long-range correlation.
Accordingly, the indifferentiable everywhere condition 
may be unified into the nonstationary condition ($\alpha \geq 2$) 
for the delocalization in the 1DDS.

An interesting question to ask is whether it is possible to 
characterize the correlation-induced localization-delocalization transition 
by  the standard one-parameter scaling theory.
The study of the finite-size scaling and critical exponents with adequate numerical
accuracy are future challenge.

\appendix

\section{Correlation function}
\label{app:correlation}
The normalized autocorrelation function $C(a,n,m)$ of the Weierstrass potential sequence $V(n)$ 
can be analytically calculated, as given for the FFM-potential by 
Petersen and Sandler \cite{petersen13}.
The explicit form becomes, 
\beq
 C(a,n,m) &=& \frac{<V(n)V(m)>}{<V(n)^2>}, \\
 &=& \frac{\displaystyle{\sum_{k=0}^{L/2}} a^{-2(2-D)k}
\cos \left[ 2\pi a^k \frac{|n-m|}{L} \right]}{\displaystyle{ \sum_{k=0}^{L/2}} a^{-2(2-D)k}},
\eeq
where $<...>$ denotes the ensemble average over the independent phases $\{ \phi_k \}$
in Eq.(\ref{eq:wei-pot}).
We set the distance between positions $\ell \equiv n-m$, 
and impose the periodic boundary conditions on the correlation function defined in 
$\ell \in [0,L/2]$.
In addition, we set $r=2\ell/L$ and $r \in [0,1]$ 
for the thermodynamic limit $L \to \infty$. 
Then the autocorrelation function can be written as 
\beq
 C(a,r) &=& \frac{\displaystyle{\sum_{k=0}^{\infty}} a^{-2(2-D)k}
\cos(\pi a^k r)}{\displaystyle{ \sum_{k=0}^{\infty}} a^{-2(2-D)k}},
\label{eq:decay}
\eeq
In the critical case, $D=3/2$, it becomes 
\beq
 C(a,r) &=& \frac{\displaystyle{\sum_{k=0}^{\infty}} a^{-k}
\cos(\pi a^k r)}{\displaystyle{ \sum_{k=0}^{\infty}} a^{-k}}.
\eeq

\begin{figure}[htbp]
\begin{center}
\includegraphics[width=8cm]{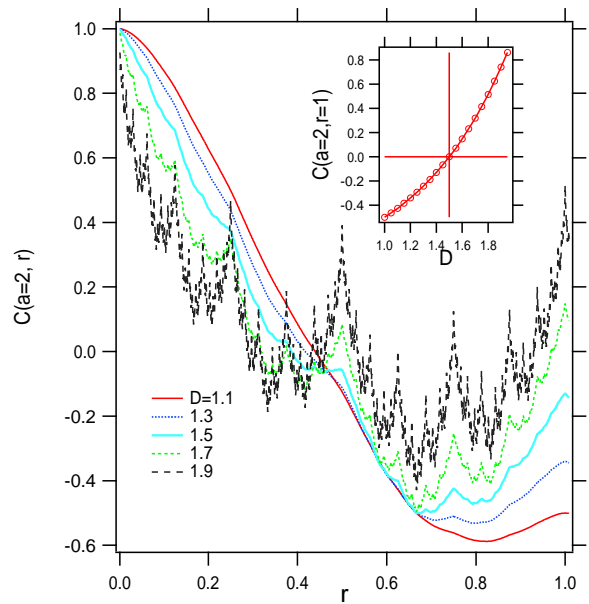}
\caption{
\label{fig:Lyap-fig1}(Color online)
The autocorrelation function $C(r)$ as a function of the distance $r$
for various fractal dimension $D$ given by 
Eq.(\ref{eq:decay}).
The parameters are $a=2$, $N=2^{15}$.
The inset shows the correlation function between the two most distant points.
}
\label{fig:corr-decay}
\end{center}
\end{figure}

Figure \ref{fig:corr-decay} shows 
the autocorrelation function $C(a=2,r)$ given by Eq.(\ref{eq:decay})
for various values of some fractal dimensions.
It follows that the correlation function rapidly decays with complex fluctuation
for $D=1.9$.
Moreover, the correlation function becomes concave for $D<3/2$ and it 
linearly decreases near $r \simeq 0$ for 
the critical value $D_c=3/2$.
The smaller the fractal dimension $D$ becomes, 
the correlation becomes more negative.
The inset of the Fig. \ref{fig:corr-decay} 
shows the values  of the  correlation function at $r=1$ as a function of the fractal dimension.
It is noted that the correlation function goes negative value at the thermodynamic limit $r=1$
for $D<3/2$.
Such a property has been 
pointed out by Petersen and Sandler in the case of the FFM-potential \cite{petersen13}.

\section*{Acknowledgments}
The author would like to thank Professor M. Goda for discussion 
about the correlation-induced delocalization at 
early stage of this study
and  Professor E.B. Starikov for proof reading of the manuscript
The author also would like to acknowledge the hospitality of 
the Physics Division of The Nippon Dental University at Niigata
for my stay, where part of this work was completed.
The sole author had responsibility for all parts of the manuscript.


\end{document}